\documentclass[aps,10pt,prx,twocolumn,superscriptaddress]{revtex4-2}

\usepackage{graphicx} 
\usepackage{comment}

\usepackage{graphicx}  
\usepackage{epstopdf}
\usepackage{dcolumn}   
\usepackage{amsmath,amssymb,dsfont}
\usepackage{wrapfig}
\usepackage{array}
\usepackage[caption=false]{subfig}
\usepackage[bookmarks=false,linkcolor=blue,urlcolor=blue,colorlinks,citecolor=blue]{hyperref}
\usepackage{exscale,relsize}
\usepackage[usenames, dvipsnames]{color}
\usepackage{bbold}
\usepackage{xcolor}
\usepackage[bottom]{footmisc}
\usepackage{makecell}
\usepackage{tabularx}
\usepackage{array}
\newcolumntype{Y}{>{\centering\arraybackslash}X}

\usepackage{cleveref}
\Crefformat{appendix}{Appendix~#2#1#3}

\usepackage{microtype}

\usepackage{siunitx}
\DeclareSIUnit{\ueV}{\micro\electronvolt}
\DeclareSIUnit{\um}{\micro\meter}
\DeclareSIUnit{\nm}{\nano\meter}
\usepackage[]{cleveref}
\Crefname{section}{Sec.}{Secs.}
\Crefname{subsection}{Sec.}{Secs.}
\Crefname{appendix}{\IfAppendix{Sec.}{Sec.}}{\IfAppendix{Secs.}{Secs.}}
\Crefname{subappendix}{\IfAppendix{Sec.}{Sec.}}{\IfAppendix{Secs.}{Secs.}}
\Crefname{equation}{Eq.}{Eqs.}
\Crefname{figure}{Fig.}{Figs.}
\Crefname{tabular}{Tab.}{Tabs.}

\newcommand*\leadauthor{\thanks{Lead authors}}

\begin{document}

\title{Excising dead components in the surface code using minimally invasive alterations:\\A performance study}

\author{Ryan~V.~Mishmash} \leadauthor \affiliation{Microsoft Quantum}
\author{Vadym~Kliuchnikov} \leadauthor \affiliation{Microsoft Quantum}
\author{Juan~Bello-Rivas} \affiliation{Microsoft Quantum}
\author{Adam Paetznick} \affiliation{Microsoft Quantum}
\author{David~Aasen} \affiliation{Microsoft Quantum}
\author{Christina~Knapp} \affiliation{Microsoft Quantum}
\author{Yue~Wu} \affiliation{Microsoft Quantum} \affiliation{Yale University}
\author{Bela~Bauer} \affiliation{Microsoft Quantum}
\author{Marcus~P.~da~Silva} \affiliation{Microsoft Quantum}
\author{Parsa~Bonderson} \affiliation{Microsoft Quantum}

\date{\today}

\begin{abstract}
The physical implementation of a large-scale error-corrected quantum processor will necessarily need to mitigate the presence of defective (thereby “dead”) physical components in its operation, for example, identified during bring-up of the device or detected in the middle of a computation. In the context of solid-state qubits, the quantum error correcting protocol operating in the presence of dead components should ideally (\emph{i}) use the same native operation set as that without dead components, (\emph{ii}) maximize salvaging of functional components, and (\emph{iii}) use a consistent global operating schedule which optimizes logical qubit performance and is compatible with the control requirements of the system. The scheme proposed by \href{https://quantum-journal.org/papers/q-2024-08-02-1429/}{Grans-Samuelsson et al. [Quantum \textbf{8}, 1429 (2024)]} satisfies all three of these criteria: it effectively excises (cuts out) dead components from the surface code using minimally invasive alterations (MIA). We conduct extensive numerical simulations of this proposal for the pairwise-measurement-based surface code protocol in the presence of dead components under circuit-level noise. To that end, we also describe techniques to automatically construct performant check (detector) bases directly from circuits without manual circuit annotation, which may be of independent interest. Both the MIA scheme and this automated check basis computation can be readily used with measurement-based as well as CNOT-based circuits, and the results presented here demonstrate state-of-the-art performance.
\end{abstract}

\maketitle

\section{Introduction}

Standard quantum error correction (QEC) protocols encode \emph{logical} qubits into the entangled states of arrays of \emph{physical} qubits and thereby make the encoded state more robust against noise on the physical qubits. A crucial assumption is that the errors that occur on the physical qubits are largely uncorrelated in space and time.
This assumption can be broken by qubits that are ``dead'', for example due to fabrication faults or inability to tune them. Such qubits can be thought of as having an error in every operation, thus leading to errors highly correlated in space and time, which
need to be handled differently, usually by modifying the circuits implementing a fault-tolerant protocol to avoid them altogether. Stace et al.~\cite{stace_error_2009} introduced a general strategy applicable to the surface code 
which involves measuring ``superstabilizers'' which avoid the dead components but still allow detection of stochastic errors in their vicinity; this approach effectively amounts to operating the surface code on a new lattice which differs from the original square lattice only locally near the dead components. Adapting this idea to concrete physical circuits implementing the surface code has been an active area of research ever since~\cite{nagayama_surface_2017, Auger, Strikis2023, Siegel2023adaptivesurfacecode, GransSamuelsson2024improvedpairwise, Lin2024, wei_low-overhead_2024, LUCI, SnL}. A particularly influential strategy, as pioneered by Auger et al.~\cite{Auger}, is to measure the superstabilizers indirectly by inferring them from non-commuting ``gauge operators'' near the dead components, whereby the protocol is operated locally as a subsystem code~\cite{poulin_stabilizer_2005, bacon_2006, bombin_topological_2010, bravyi_subsystem_2013, higgott_subsystem_2021}. In particular, the Auger scheme relies only on the same native operation set as the parent (dead-component-free) surface code protocol (e.g., it does not require additional SWAP gates~\cite{nagayama_surface_2017}) and exhibits a competitive performance baseline.

\begin{figure*}[t]
    \centering
    \includegraphics[width=0.9\linewidth]{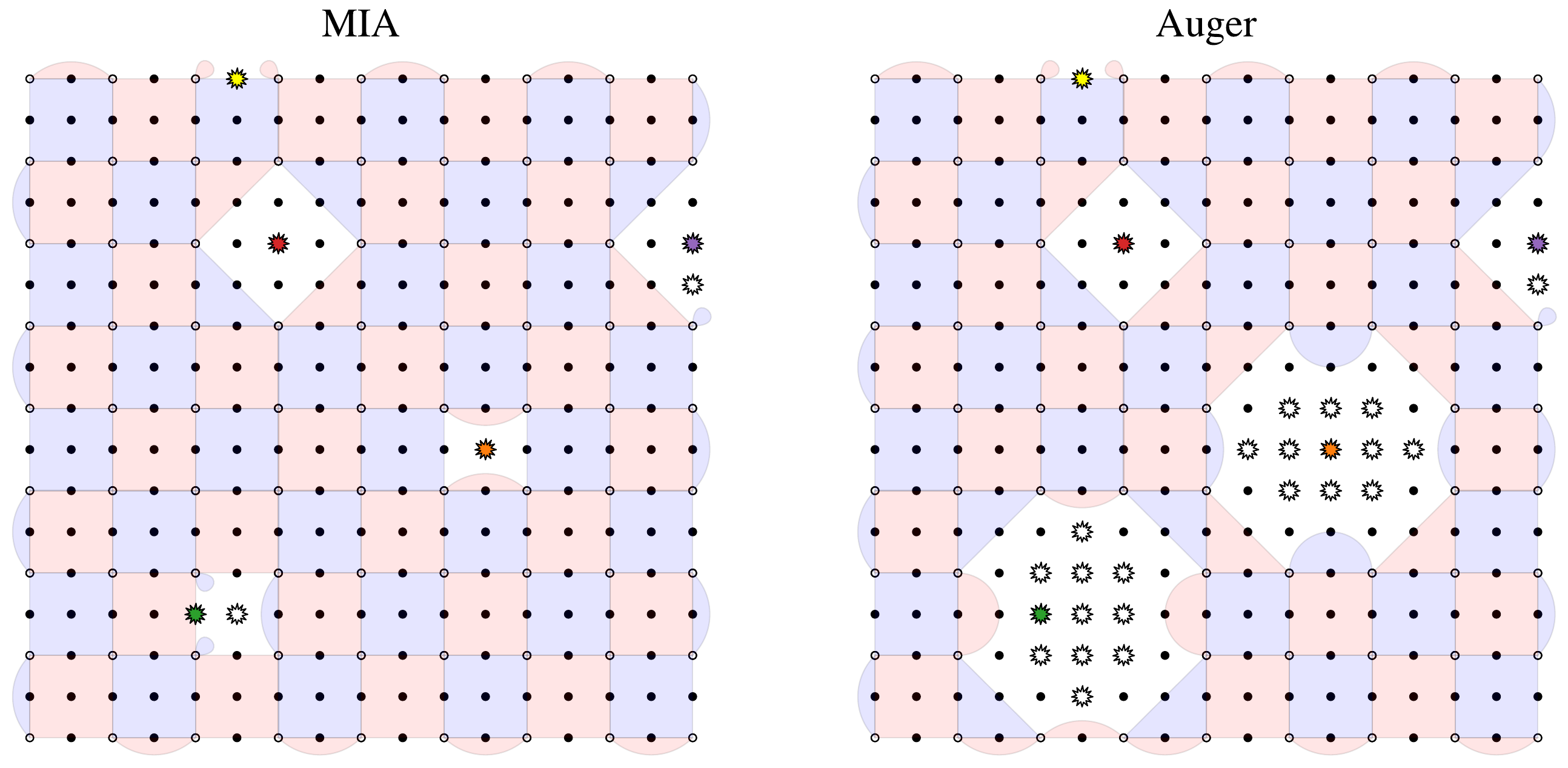}
    \caption{
    $n$-gons configurations representing circuits for a 3aux measurement-based surface code under the MIA (left) and Auger (right) schemes on a $L=9$ square patch with ``good'' (hook-benign) boundary conditions for a dead-qubit configuration consisting of a single dead qubit of each of the following types: bulk data qubit (red), bulk A/C auxiliary qubit (green), bulk B auxiliary qubit (orange), boundary data qubit (purple), and boundary 2-gon auxiliary qubit (yellow). Open (filled) circles represent data (auxiliary) qubits; and $X$($Z$)-type $n$-gons are colored red (blue). For our ``Auger'' simulations in this work (and the circuit represented above), dead auxiliary qubits in 4-gons result in the collateral disabling of that 4-gon's data qubits, while, for simplicity, dead qubits in the boundary 2-gons are treated the same in each scheme (see Appendix~\ref{app:boundary} for a more in-depth discussion of our boundary treatment).
    }
    \label{fig:ngon_overlays}
\end{figure*}

Here we study in depth via detailed performance simulations a new strategy---first described in Ref.~\onlinecite{GransSamuelsson2024improvedpairwise} and here referred to as the minimally invasive alterations (MIA) scheme---that leads to several key improvements over Auger without sacrificing simplicity. Specifically, the MIA scheme---in addition to using the same native operation set as the parent protocol---(\emph{i}) maximizes salvaging of functional components of the system, thereby minimizing the damage to the code in terms of the number and size of superplaquettes; (\emph{ii}) uses a scheduling strategy which measures both the original plaquette stabilizers and the superstabilizers 
without deviating from the global schedule of the parent protocol; and (\emph{iii}), as an added bonus, requires no {\em ad hoc} modifications to the circuit near the boundary of the code. Importantly, since the MIA scheme does not change the global schedule of the parent dead-component-free circuit, it has particularly friendly requirements from the perspective of large-scale digital control. Central to our scheme is the representation of the surface code as a collection of ``$n$-gons'', i.e., circuits which measure Pauli stabilizers of $n$ specified data qubits. The full syndrome extraction circuit implementing the surface code can then be viewed as a tiling in space-time of these $n$-gon units, and the MIA scheme arises naturally from a set of rules to simply \emph{reduce} and minimally \emph{split} the $n$-gons given a configuration of dead components~\cite{GransSamuelsson2024improvedpairwise}---the above desirable features then follow directly.

In this paper, we focus our simulations and analysis on the pairwise-measurement-based surface code realization of Ref.~\onlinecite{GransSamuelsson2024improvedpairwise} (which we nickname {\em 3aux}~\footnote{In this scheme, each bulk (4-gon) surface code stabilizer employs three auxiliary qubits in its stabilizer extraction circuit, hence the name.}) as it is very well-suited toward Majorana-based hardware~\cite{aasen_roadmap_2025}; however, we stress that the MIA scheme itself can be applied universally to any surface code realization (and potentially other quantum codes), including the prototypical CNOT-based one.

The circuit modification in the MIA scheme must be accompanied by modifications to the decoder. We describe new details of constructing decoders directly from the modified circuits using matroid techniques as described in Beverland et al.~\cite{beverland2024faulttolerancestabilizerchannels}. Specifically, for each modified circuit we compute checks that lead to a decoding graph for a minimal one-qubit Pauli noise model and use these checks for decoding full circuit-level noise in the 3aux protocol. We use this pipeline to efficiently explore a large number of dead qubit configurations in the context of an independent and identically distributed (IID) qubit failure model to characterize the performance of our scheme and ultimately inform physical qubit yield requirements given a logical performance requirement on the constructed logical qubit.

We directly compare the performance of the MIA scheme to that of the Auger scheme as applied to 3aux, taking an efficient pipelined schedule for syndrome extraction in each case. Indeed we find a very definitive performance improvement comparable to that exhibited by the recent schemes described in Refs.~\cite{LUCI, SnL}.

\section{MIA scheme} \label{sec:MIA}

Here we review the MIA scheme for dead-component handling as first introduced by Grans-Samuelsson et al.~\footnote{See Sec.~5 of Ref.~\onlinecite{GransSamuelsson2024improvedpairwise}, noting that in this reference the scheme was not as yet termed \emph{MIA}.}. The key conceptual feature of the scheme is to view the dead-component-free surface code for a given physical realization as a collection of so-called $n$-gon units, where each $n$-gon is a physical implementation of a syndrome extraction circuit for $n$ specific data qubits. For example, a patch of the rotated surface code implementing a quantum memory is a collection of 4-gons (for the bulk stabilizers) and 2-gons (for the boundary stabilizers) stitched together in space to form a single logical qubit. The circuits associated with the $n$-gons depend on the physical realization of the code as well as scheme and schedule used for syndrome extraction. For the prototypical surface code implementation using CNOT entangling gates and single-qubit measurements, each 4- and 2-gon contains the single auxiliary qubit for syndrome measurement (as well as the data qubits whose syndrome is getting measured). For the 3aux implementation using pairwise two-qubit measurements, there is likewise a library of circuits for the $n$-gons which are designed to give the surface code stabilizers as measured observables~\cite{kliuchnikov2023stabver} (see Ref.~\onlinecite{GransSamuelsson2024improvedpairwise} for details).

For a dead-component-free surface code, the $n$-gons of course merely represent the (local) circuits for extracting the code's stabilizers; however, this framework is particularly convenient for implementing circuits to extract \emph{superstabilizers} in the presence of dead components by operating the system locally as a subsystem code: given a collection of $n$-gons representing a surface code protocol and a configuration of dead components, the MIA scheme proceeds by simply \emph{reducing} and \emph{splitting} the $n$-gons according to a set of prescribed rules. $n$-gon reduction is associated with death of data qubits in the corresponding circuit, while $n$-gon splitting is associated with dead auxiliary qubits and non-functioning connections (e.g., couplers) between qubits; see Figs.~10-12 and 30-32 of Ref.~\onlinecite{GransSamuelsson2024improvedpairwise} for details. The key point is that the resulting minimally reduced/split $n$-gons near the dead components serve as ``gauge operator'' measurements to infer superstabilizers in a particularly efficient and simple way.  For the case of dead data qubits, the MIA procedure and that of Auger~\cite{Auger} are identical regardless of physical realization. However, the two schemes differ drastically in how dead auxiliary qubits and connections are handled, with the MIA being significantly more efficient in terms of component salvaging: for Auger, any single dead auxiliary qubit or connection in a bulk 4-gon leads to the removal of the entire plaquette, while the MIA's tactical $n$-gon splitting procedure causes much less damage to the code. We demonstrate these differences in Fig.~\ref{fig:ngon_overlays} for the case of the 3aux implementation. Note that the result of such Auger-like component salvaging also leads to a description of the final circuit as a collection of $n$-gons, albeit with significantly larger holes and thus damage to the code.

In addition to minimizing code damage \emph{in space}, the MIA scheme also features notable improvements over the Auger scheme---and in fact many other dead-component mitigation schemes---with respect to circuit scheduling \emph{in time}. For the 3aux implementation, it is very natural to use a \emph{pipelined} syndrome extraction schedule such that both $X$- and $Z$-type stabilizers are effectively both measured in every round of syndrome extraction (four physical cycles for the most compact circuit). Such a scheduling strategy is in fact naturally suitable with subsystem-code-based protocols for dead component mitigation: one does not need to measure $X$- and $Z$-type damaged plaquette stabilizers in alternating rounds since the pipelined schedule effectively alternates $X$ and $Z$ \emph{within each round}. However, as was pointed out in Ref.~\onlinecite{GransSamuelsson2024improvedpairwise}, dead-component protocols can also be operated on an \emph{interleaved} schedule---as typically utilized with superconducting qubits---in such a way that strict alternation is not actually necessary: with an appropriate scheduling, the superstabilizer measurements for damaged plaquettes instead build up over \emph{several rounds} of syndrome extraction, the number of which depends on the configuration of dead components. Because the simulations performed in this paper focus on the 3aux surface code implementation for which pipelining is most natural, we only consider maximally efficient pipelined schedules using both MIA and Auger-like component salvaging procedures.

One final advantage of the MIA scheme concerns its treatment of dead components near the patch boundary. For existing dead-component schemes, e.g, those in Refs.~\cite{Auger,LUCI,SnL}, it is typical to \emph{deform} the boundary to accommodate nearby dead components and operate the protocol similarly to the bulk. However, the above features of the MIA scheme---circuits derived from an optimal $n$-gon reduction/splitting procedure and lack of the need for measuring $X$- and $Z$-type damaged plaquette stabilizers in alternating rounds (for either pipelined or interleaved schedules)---allow us to treat the boundary on exactly the same footing as the bulk without any \emph{ad hoc} deformation: the code heals up naturally with only a modest hit to distance/performance (see Appendix~\ref{app:boundary} for details).

In Sec.~\ref{sec:performance}, we present extensive large-scale Monte Carlo simulations of both the MIA and Auger schemes in the context of the 3aux surface code implementation using a compact pipelined schedule and show a large performance separation between the two schemes---comparable that shown for recently introduced advanced proposals in the context of CNOT-based implementations~\cite{LUCI,SnL}. In addition to its performance, perhaps the most overall desirable quality of the MIA scheme is its simplicity: the minimally reduced/split $n$-gons use the same native operation set and run on the same global operating schedule as the parent surface code protocol. This clearly has advantages if operating the system using large-scale (cryo) digital control, as is central to the measurement-based Majorana architecture~\cite{aasen_roadmap_2025}.

\section{Decoder construction}

Decoder construction for circuits with dead components presents unique challenges, as standard approaches typically require manual annotation via somewhat \emph{ad hoc} logic. We present a fully automated approach that constructs performant decoders directly from modified circuits without any manual input, enabling efficient exploration of many dead-component configurations.

\subsection{Overview and setup} \label{sec:decoder_overview}

In our numerical experiments, we use the following setup: 
a noiseless prior circuit, a noisy circuit of a prescribed number of syndrome extraction rounds, and a noiseless posterior circuit.
The prior circuit is equivalent to a circuit that measures stabilizers of an input code;
similarly, the posterior circuit measures stabilizers of the output code.
The prior and posterior circuits specify the input and output codes of the noisy circuit. 
Using this circuit triple, we compute a basis of circuit (or space-time) checks~\footnote{These are also commonly referred to as {\em detectors}~\cite{Gidney2021stimfaststabilizer, Higgott2025sparseblossom}.}, logical effect, and check matrices~\cite{beverland2024faulttolerancestabilizerchannels} without any further user input. 
Importantly, we automatically find a suitable check basis that enables decoding using hyperedge splitting~\cite{delfosse2023splittingdecoderscorrectinghypergraph} (not to be confused with the $n$-gon splitting) and PyMatching~\cite{higgott2021pymatching, Higgott2025sparseblossom}.

\subsection{Logical effect and check matrices from circuits}

We obtain all data needed for decoder construction as follows.
Recall that checks are parities of measurement outcomes that have a fixed, known value in the absence of circuit faults.
To find a basis of checks, we compute the outcome code~\cite{delfosse2023spacetime} of the concatenated circuit triple using the outcome-complete simulation algorithm described in Ref.~\onlinecite{kliuchnikov2023stabver}.

The logical effect and check matrices have the following structure: the columns of both matrices correspond to elementary circuit faults.
The rows of the check matrix correspond to circuit checks, 
while the rows of the logical effect matrix correspond to logical measurement outcomes of the noisy circuit and observables of the output code. 
Undetectable bad faults correspond to binary vectors in the null space of the check matrix but outside the null space of the logical effect matrix---in other words, bad faults do not violate any checks but have non-trivial logical action.

We compute the check matrix using standard fault propagation techniques.
The input and output codes are determined from the prior and posterior circuits using the circuit general form algorithm from Ref.~\onlinecite{kliuchnikov2023stabver}.
We then compute the logical effect matrix using the logical action algorithm of Ref.~\onlinecite{kliuchnikov2023stabver}, given the input and output codes of the noisy circuit.
In summary, we compute both logical effect and check matrices starting from the circuit triple without requiring additional input beyond the circuit specifications.
In other words, our approach does not require manual annotations of measurement outcomes.

\subsection{Graph realization and check basis optimization} \label{sec:graph_realization}

The main challenge of this fully automated approach is that the check basis found by the outcome code does not lead to a \emph{graph-like} check matrix (i.e., a matrix with column weights of at most two).
Similarly, it may not produce a check matrix suitable for heuristics that reduce hypergraph decoding (matrices with column weights above two) to graph-like decoding, 
such as hyperedge splitting~\cite{delfosse2023splittingdecoderscorrectinghypergraph}. 

To address this challenge, we use a graph realization algorithm. 
Changing the check basis corresponds to left-multiplying the check matrix by an invertible matrix, which we refer to as the basis change matrix.
A matrix for which there exists a basis change that transforms it into a graph-like matrix is called a \emph{graphic matrix}. 
Efficient algorithms exist to determine whether a matrix is graphic and to find the corresponding basis transformation~\cite{bixbywagner, vanderhulst2025rowgraphrealization}. 
This is known in the literature as the \emph{graph realization problem}.

\begin{table}[h!]
\centering
\begin{tabularx}{\columnwidth}{c|Y|Y||Y|Y}
 & \multicolumn{2}{c||}{$L=7$, 504 configs.} & \multicolumn{2}{c}{$L=9$, 509 configs.} \\
\cline{2-5}
 & MIA & Auger & MIA & Auger \\
 \hline\hline
No logical qubit & 0 & 10 & 0 & 7 \\
\hline
\makecell{Non-graphic \\ with logical qubit} & 0 & 0 & 1 & 2 \\
\hline
\multicolumn{1}{c|}{$n_\mathrm{configs}^\mathrm{total}$} & \multicolumn{2}{c||}{504} & \multicolumn{2}{c}{506} \\
\hline
\end{tabularx}
\caption{
Summary of dead-qubit configuration numbers considered for simulation in Sec.~\ref{sec:performance} with respect to whether or not a valid surface code logical qubit is formed and, given that a logical qubit formed, whether or not the algorithm in Sec.~\ref{sec:graph_realization} produces a graphic check matrix for the minimal elementary fault set described therin. (For the latter, we only count cases for which a logical qubit is formed as otherwise the circuit may be pathological---lack of a logical qubit is due to percolation of dead and/or disabled components across the lattice~\cite{Auger}.) For example, for the $L=9$ system, we considered 509 total dead-qubit configurations; all of them formed a logical qubit under MIA, while 7 failed to do so under Auger; of those configurations that did form a logical qubit, we were able to construct graphic check matrices for all but 1 (2) of them under MIA (Auger). We use the $n_\mathrm{configs}^\mathrm{total}$ graphic cases under both schemes for the data presented in Figs.~\ref{fig:many_dead},\,\ref{fig:histograms}, and \ref{fig:hgUF_comp} below.
}
\label{tab:graphicness_summary}
\end{table}

We use graph realization algorithms to find a check basis and corresponding check matrix that enables decoding via hyperedge splitting of the full noise model for many dead-component configurations. 
Our strategy begins by considering the check matrix for a minimal subset of elementary faults that produces a matrix of the same rank as the full circuit-level noise check matrix. 
This minimal subset consists of single-qubit $X$ and $Z$ faults on all input qubits of the noisy circuit, plus single-qubit $X$ and $Z$ faults immediately following each and every operation in the noisy circuit.
We find that in most cases (see Table~\ref{tab:graphicness_summary} and next subsection), the check matrix for this minimal noise model is graphic, allowing us to compute a suitable check basis. 
We then use this basis to decode the full circuit-level noise model with hyperedge splitting~\cite{delfosse2023splittingdecoderscorrectinghypergraph} and PyMatching (uncorrelated minimum-weight perfect matching)~\cite{higgott2021pymatching,Higgott2025sparseblossom}, which we refer to below as ``splitting + PyMatching''.

\subsection{Implementation details and algorithm performance}

We discuss practical aspects of implementing the graphicness algorithm and computational considerations. 
We developed a Rust port of a compact Java implementation~\cite{grp-impl} of the Bixby-Wagner algorithm~\cite{bixbywagner}. 
The algorithm runs in almost linear time in the number of non-zero entries of the row-reduced echelon form of the check matrix. 
The main computational requirement is computing the row-reduced echelon form of the check matrix, which is a well-studied problem for binary matrices. 
Practical algorithms can handle matrices of size $10{,}000 \times 10{,}000$ in a few seconds~\cite{m4ri}.

When applying our techniques to standard surface code circuits, we can obtain graphs that differ from those typically constructed manually.
This occurs because multiple solutions exist for the graph realization problem. 
Any two solutions to the same problem are two-isomorphic to each other~\cite{twoisomorphic}; two-isomorphism can be understood in terms of elementary graph transformations~\cite{twoisomorphic} reminiscent of local complementation in graph states. 
The specific solution we find depends on the column ordering of the check matrix. 
Random column permutations can generate multiple solutions, and algorithms exist for enumerating all graphs two-isomorphic to a given graph~\cite{twoisomorphismcount}.

For the decoding simulations presented in Sec.~\ref{sec:performance}, we use the unmodified check basis produced by our implementation and---after hyperedge splitting of the full circuit-level noise model---find that it produces the expected logical performance (e.g., we can exactly reproduce the dead-component-free performance results from Ref.~\onlinecite{GransSamuelsson2024improvedpairwise}, which were obtained therin with a manual plaquette-labeling strategy). In our study of the component failure model described in Sec.~\ref{sec:failure_model}, where we consider $\sim$$2{,}000$ 3aux circuits for the MIA and Auger schemes, there are $\mathcal{O}(1)$ cases for which the above algorithm fails to produce a graphic check matrix for the minimal noise model (see Table~\ref{tab:graphicness_summary} for summary statistics). We simply throw out those few cases from the representative populations used for the results in Figs.~\ref{fig:many_dead},\,\ref{fig:histograms}, and \ref{fig:hgUF_comp}
\footnote{Note that the multiple solutions to the graph realization problem, as discussed in the previous paragraph, can prove useful for heuristics for handling non-graphic cases.}.

\section{Performance simulations} \label{sec:performance}

In this section, we present our main simulation results, applying the above automated check determination and decoder construction machinery to run logical performance estimation of the MIA and Auger circuits for dead-component-ridden 3aux surface code circuits.

\subsection{Circuit, noise model, and simulation details}

For our simulations, we take the simplest baseline 3aux memory protocol from Ref.~\onlinecite{GransSamuelsson2024improvedpairwise}. For concreteness, we make the ``good'' choice of boundary orientation. This choice makes hook errors benign for the dead-component-free case such that a square patch with $L$ data qubits on a side produces a logical qubit with fault distance \cite{bombin_logical_2023,beverland2024faulttolerancestabilizerchannels} $d_f = L$ using $n_\mathrm{qubits} = (2L-1)^2$ total qubits. Furthermore, we use the most compact (4-step period) measurement circuit compatible with a double-rail semiconductor layout in Majorana hardware. As mentioned above, we only consider pipelined schedules for the $n$-gon circuits. See Figs.~4 and 8 of Ref.~\onlinecite{GransSamuelsson2024improvedpairwise} for the specific circuits for all requisite $n$-gons.

For our ``\emph{Auger}'' simulations, we implement a component salvaging strategy which reduces to that of Ref.~\onlinecite{Auger} for the \emph{bulk} of a 3aux surface code patch (all 4-gons), while for simplicity we use the MIA salvaging procedure for dead qubits involved in boundary 2-gons (i.e., boundary data qubits and 2-gon auxiliary qubits)~\footnote{We expect using the Auger boundary deformation strategy would only lead to further relative performance degradation for Auger, but we have not studied this in detail.}. This amounts to collaterally disabling all data qubits of a bulk 4-gon if \emph{any} auxiliary qubits of the $4$-gon are dead. The circuits thereby resulting from the MIA procedure and the Auger procedure can each ultimately be described by a collection of $n$-gons---see Fig.~\ref{fig:ngon_overlays} for the $n$-gon representations for circuits given various characteristic qubit failures for each scheme.

\begin{figure*}
    \centering
    \includegraphics[width=1.0\linewidth]{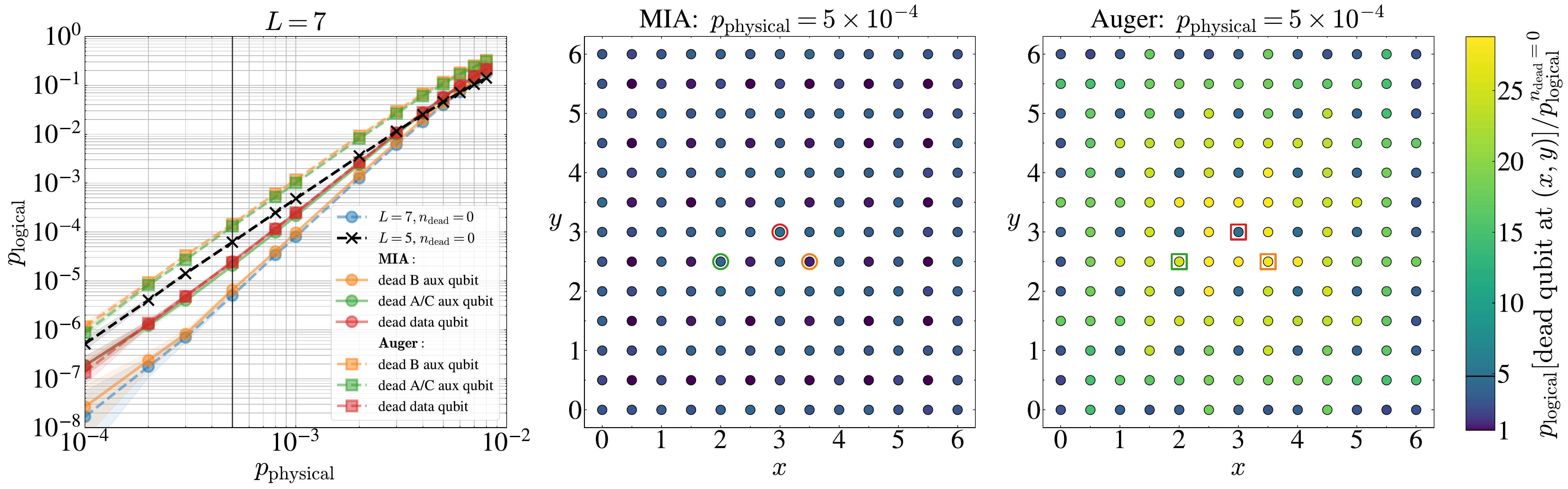}
    \caption{
    Logical performance of 3aux circuits with a single dead qubit ($n_\mathrm{dead}=1$). (left) We plot $p_\mathrm{logical}$ (obtained using splitting + PyMatching) as a function of $p_\mathrm{physical}$ for circuits obtained from the MIA and Auger schemes for three dead-qubit configurations in the $L=7$ system, each consisting of a single dead data (red), A/C auxiliary (green), and B auxiliary (green) qubit near the middle of the patch (the specific selected dead qubit for each case is highlighted in the middle and right panels). The logical performance improvement using the MIA (circle symbols) versus Auger (square symbols) is very pronounced for dead auxiliary qubits (green and orange), while the two schemes produce identical circuits and performance for a single dead data qubit (red). The black dashed line corresponds to $L=7-2=5$ with $n_\mathrm{dead}=0$; for $L=7$ and $n_\mathrm{dead}=1$, the MIA data always beats this reference, while Auger lies above it for single auxiliary qubit death. Fixing $p_\mathrm{physical} = 5 \times 10^{-4}$ (vertical line in left panel), we plot the logical error rate obtained when a single qubit at position $(x,y)$ is dead divided by the dead-component-free ($n_\mathrm{dead}=0$) logical error rate---for all qubits in the patch---under the MIA (middle) and Auger (right) schemes; the maximum relative logical error rate observed in the MIA scheme is marked with the horizontal line in the color bar. In summary, the MIA scheme handles single dead B auxiliary qubits with almost no performance degradation (darkest points in middle panel), with single dead data and A/C auxiliary qubits giving similar performance; on the other hand, single dead auxiliary qubits under Auger are extremely costly.
    }
    \label{fig:single_dead}
\end{figure*}

Because all stabilizers---$X$- and $Z$-type damaged and undamaged stabilizers---are effectively measured in every round given our pipelined schedule of $n$-gons, we simulate $L$ noisy rounds of syndrome extraction for the both the MIA and Auger schemes; these $L$ noisy rounds are buttressed by two noiseless rounds before and after the memory gate (to ensure that the circuit measures all stabilizers of the modified surface code---see Sec.~\ref{sec:decoder_overview}). We use the standard depolarizing-like (``EM3'') noise model for measurement-based circuits~\cite{chao_optimization_2020,gidney_fault-tolerant_2021,GransSamuelsson2024improvedpairwise} parameterized by physical error rate $p_\mathrm{physical}$. Calculated logical error rates, $p_\mathrm{logical}$, correspond to incorrect recovery of \emph{any} logical observable, i.e., obtained logical Pauli error + correction different from the identity under this setting. Error bars for individual $p_\mathrm{logical}$ runs correspond to 95\% credible intervals of the posterior beta distribution given the observed number of logical failures and completed trials, assuming a uniform prior distribution for the logical error rate. We target simulating up to $\sim$$10^{8}$ trials or $\sim$$10^{4}$ failures, whichever comes first. 

We set up decoders using the graph-realization-based automated check basis construction described in Sec.~\ref{sec:graph_realization} in which we construct a graphic check matrix from a minimal elementary noise model which is a subset of the full circuit-level noise model. Since the latter is not graphic with respect to this check basis, we must either attempt to split (pre-decode) the hyperedges in the full noise model
~\cite{delfosse2023splittingdecoderscorrectinghypergraph} and subsequently use a graph-based decoder, or employ a more general hypergraph-based decoder. In what follows, we mostly employ the former strategy for studying dead-component-ridden circuits: the data in Figs.~\ref{fig:single_dead} and \ref{fig:many_dead} was obtained using hyperedge splitting + PyMatching (see Ref.~\onlinecite{GransSamuelsson2024improvedpairwise} for a review of how such a ``splitting decoder'' is constructed in this context). In Fig.~\ref{fig:hgUF_comp}, we investigate the performance of a hypergraph-based decoder, namely an implementation~\cite{mwpf2025} of weighted hypergraph union-find~\cite{delfosse_toward_2021}.

\subsection{Component failure model and configuration sampling} \label{sec:failure_model}

Dead Majorana qubits could occur for several reasons; for example, individual topological wire segments could not exhibit a topological phase, which would lead to a single qubit failing, or quantum dots used to perform single- and two-qubit measurements could not exhibit a favorable tuning regime, in which case certain qubit operations need to be disabled. While the MIA scheme can handle both of these cases, we focus here on uncorrelated single-qubit failures, as would arise, e.g., from wire segments failing. That is, we consider a component failure model in which each qubit in a patch fails independently with probability $p_\mathrm{failure}$.

In order to efficiently assess logical performance over a range of $p_\mathrm{failure}$ without having to generate independent samples for a particular value thereof, we employ the following sampling strategy. First, for each considered patch size, we select $n_\mathrm{configs}^\mathrm{total}\simeq500$ unique dead qubit configurations to represent the entire population of configurations. To build these representative populations, for each $k = 0,1,\dots,k_\mathrm{max}$ total dead qubits, we randomly select $n_k$ unique configurations such that $\sum_{k=0}^\mathrm{k_\mathrm{max}} n_k = n_\mathrm{configs}^\mathrm{total}$; $k_\mathrm{max}$ and the $n_k$ are chosen to capture typical configurations for $0 \leq p_\mathrm{failure} \leq 2\%$. We then carry out full performance simulations for these selected configurations. Finally, to approximate sampling from the full population for a given value of $p_\mathrm{failure}$, we choose $n_\mathrm{configs}$ configurations with replacement from these $n_\mathrm{configs}^\mathrm{total}$ configurations as follows: for each sample, we first select a $k$ according to the binomial distribution $p(k; n_\mathrm{qubits}, p_\mathrm{failure}) = \binom{n_\mathrm{qubits}}{k} p_\mathrm{failure}^k (1 - p_\mathrm{failure})^{n_\mathrm{qubits}-k}$, and then for this selected $k$ we randomly draw a configuration uniformly from the subpopulation of $n_\mathrm{dead} = k$ dead qubits of size $n_k$; this is repeated $n_\mathrm{configs}$ times. It is these $n_\mathrm{configs}$ configurations for which we report a distribution of performance data for a fixed value of $p_\mathrm{failure}$, where below we choose $n_\mathrm{configs} = 200$.

\subsection{Results}

We begin by analyzing dead-component configurations consisting of a \emph{single dead qubit} in the 3aux surface code. For the memory experiment we consider, there are three main classes of qubits: data qubits, ``A/C'' auxiliary qubits, and ``B'' auxiliary qubits~\footnote{These auxiliary qubit labels correspond to the labels used in Ref.~\onlinecite{GransSamuelsson2024improvedpairwise}.}. When placed on a square grid such that the data qubits have integer coordinates in both the $x$ and $y$ directions, A/C auxiliary qubits live on the edge of bulk 4-gons and have one integer coordinate and one half-integer coordinate; while B auxiliary qubits live at the center of bulk 4-gons with both coordinates being half-integer. The $Z/X$-type $n$-gons are drawn in blue/red in Fig.~\ref{fig:ngon_overlays}. The memory patches we simulate have logical operators as follows: the $Z$ ($X$) logical observable, i.e., product of Pauli-$Z$($X$) operators on data qubits, traverses horizontally (vertically). Analyzing the instantaneous stabilizer group for the case of a single dead qubit in this patch, we see that under the MIA strategy the \emph{code distance} is altered as follows: reduced by one for both logical observables for a single dead data qubit, reduced by one for only one of the logical observables for a single dead A/C auxiliary qubit, and unchanged given a single dead B auxiliary qubit. This logic also holds for dead qubits on the boundary (auxiliary qubits on the boundary are purely of ``A/C'' type for the considered boundaries). As mentioned above, the MIA and Auger schemes coincide for the case of only data qubit death; however, the Auger scheme leads to a reduction of code distance by two for both logical observables if \emph{any} auxiliary qubit in a bulk 4-gon is dead---this bodes particularly poorly for directly applying the original Auger scheme to 3aux.

In Fig.~\ref{fig:single_dead}, we show performance data on an $L=7$ (= target fault distance $d_f$) 3aux patch corresponding to a single dead data qubit (red), a single dead A/C auxiliary qubit (green), and a single dead B auxiliary qubit (orange) for both the MIA (circle) and Auger (square) schemes. In the left panel, we show full performance curves; the case of no dead qubits ($L=7, n_\mathrm{dead}=0$) is also shown, as is the curve for a dead-component-free patch at one patch size smaller ($L=5, n_\mathrm{dead}=0$)~\footnote{The data for the $n_\mathrm{dead}=0$ circuits are shifted up slightly from those in Ref.~\onlinecite{GransSamuelsson2024improvedpairwise} as the noise model implemented in code in Ref.~\onlinecite{GransSamuelsson2024improvedpairwise} neglected to include noise on idle locations for these circuits. While there is no idling in the bulk of the lattice, idle locations are present for the boundary qubits.}. The precise qubits which are dead in the simulations in the left panel are marked in the middle and right panels. These results confirm the intuition from the stabilizer group evolution discussed above. In particular, the MIA scheme handles any single dead qubit in the patch such that it still comfortably beats the dead-component-free $L=5$ performance, while applying Auger to configurations with dead auxiliary qubits lead to worse performance than even the dead-component-free $L=5$ system. In the middle and right panels of Fig.~\ref{fig:single_dead}, we show data at a fixed $p_\mathrm{physical} = 5 \times 10^{-4}$: every point in the lattice represents performance relative to the dead-component-free case (at $L=7$) if the qubit residing at that position were dead and handled by either the MIA (middle) or Auger (right). In the case of MIA, performance is best for dead B auxiliary qubits---the corresponding ``sublattice'' structure is apparent in the data; in the case of Auger, data qubit death is most favorable (where the schemes coincide), while auxiliary qubit death is extremely costly.

\begin{figure}
    \centering
    \includegraphics[width=1.0\linewidth]{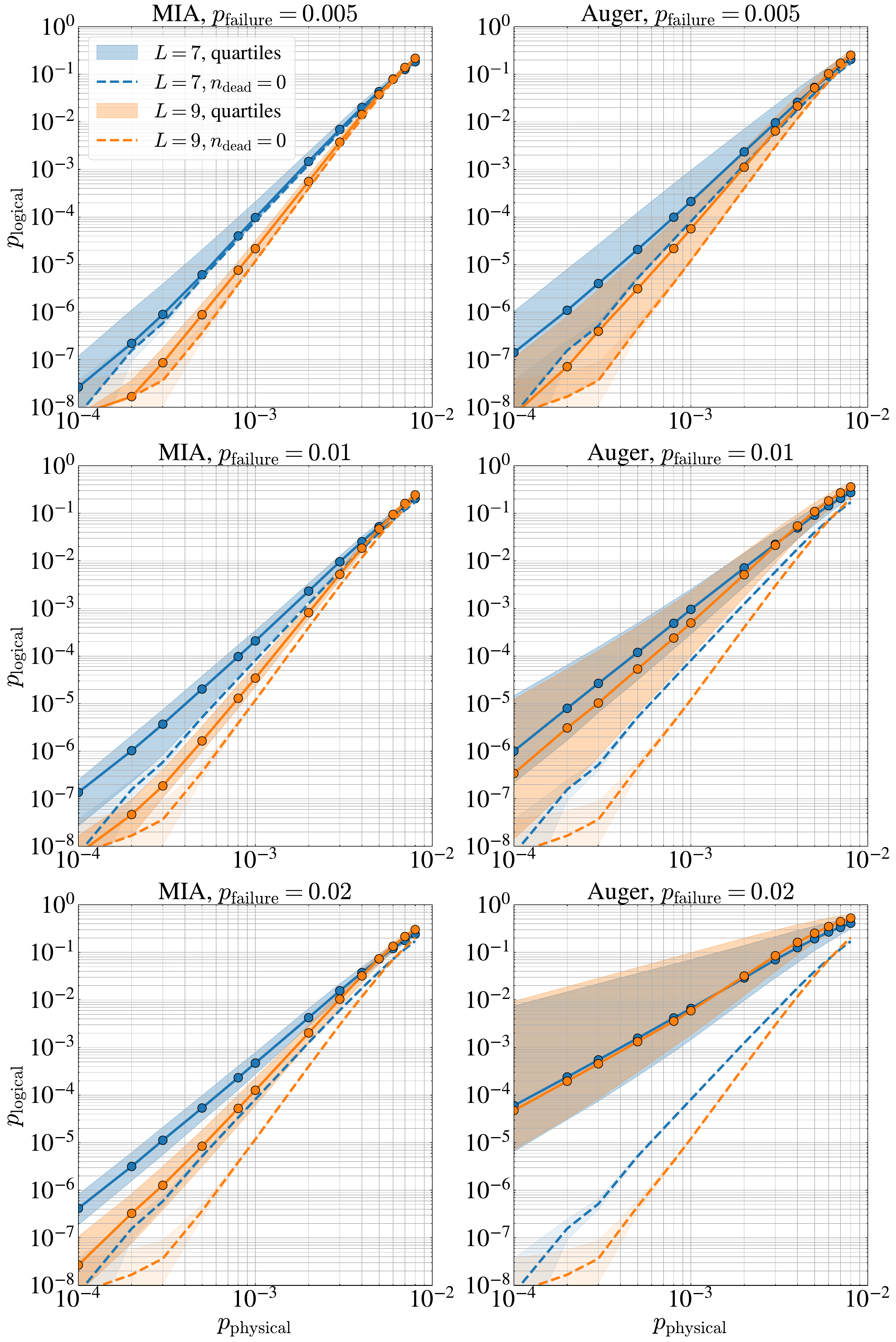}
    \caption{
    Logical performance of 3aux circuits under independent single-qubit failure model: dependence on $p_\mathrm{failure}$ and $p_\mathrm{physical}$. We sample $n_\mathrm{configs} = 200$ dead-qubit configurations with replacement from representative populations of the $L=7$ and 9 systems, of size $n_\mathrm{configs}^\mathrm{total} = 504$ and 509, respectively (see text for details). Logical performance results (obtained using splitting + PyMatching) under the MIA (left column) and Auger (right column) component salvaging strategies are shown, from top to bottom, for qubit failure rates of $p_\mathrm{failure} = 0.5\%, 1\%$, and 2\%. The dark shaded regions represent the interquartile range of the distribution of simulated $p_\mathrm{logical}$ estimates, while the solid curves and circles represent the medians. For reference, we also show the performance for the dead-component-free case ($n_\mathrm{dead} = 0$) with dashed curves; the statistical error (see text for details) is indicated by the light shaded regions. Configurations for which Auger fails to form a logical qubit (see Table~\ref{tab:graphicness_summary}) are reported here and in Fig.~\ref{fig:histograms} as $p_\mathrm{logical}^\mathrm{Auger} = 50\%.$
    }
    \label{fig:many_dead}
\end{figure}

\begin{figure*}
    \centering
    \includegraphics[width=1.0\linewidth]{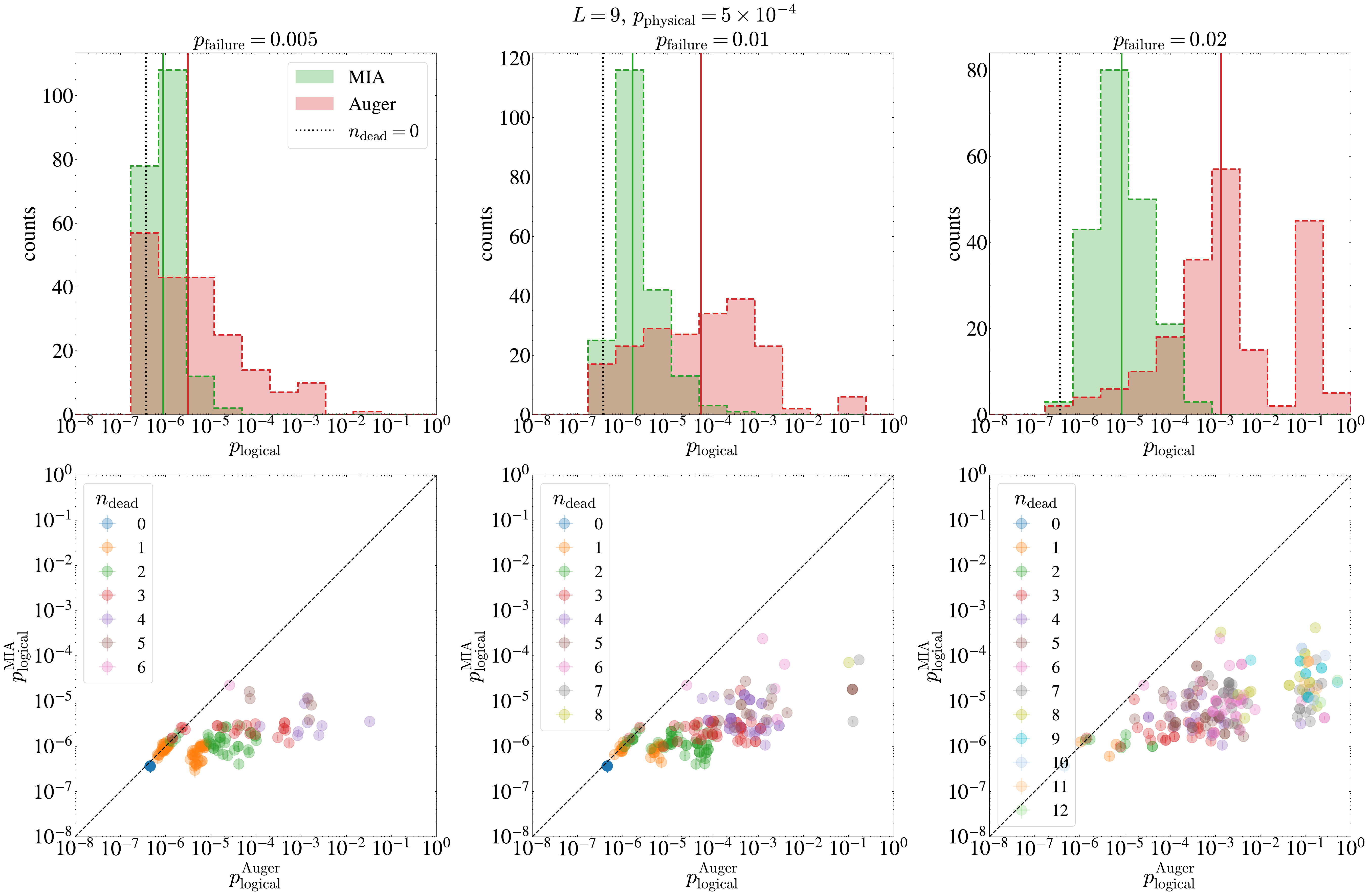}
    \caption{
    Logical performance of 3aux circuits under independent single-qubit failure model: details of distributions at fixed $p_\mathrm{physical}$. For the $L=9$ system at $p_\mathrm{physical} = 5 \times 10^{-4}$, we show (top) histograms and (bottom) scatter plots of simulated logical error rates $p_\mathrm{logical}$ under the MIA (green in top row, vertical axis in bottom row) and Auger (red in top row, horizontal axis in bottom row) schemes. In each histogram plot, we indicate the median of the respective distributions with solid vertical lines and the dead-component-free $p_\mathrm{logical}$ with a vertical dotted line ($n_\mathrm{dead}=0$). In the scatter plots, we color each point according to the total number of dead qubits $n_\mathrm{dead}$ in the selected configuration; vertical and horizontal error bars (barely visible) denote the statistical uncertainty of the respective estimates of $p_\mathrm{logical}$. Configurations for which Auger did not produce a logical qubit are reported here and in Fig.~\ref{fig:many_dead} as $p_\mathrm{logical}^\mathrm{Auger} = 0.5$. There exist configurations with a four-to-five order of magnitude separation in $p_\mathrm{logical}$ between the two schemes.
    }
    \label{fig:histograms}
\end{figure*}

Next we turn to a study of the component failure model described above in which each qubit fails independently with probability $p_\mathrm{failure}$. In Fig.~\ref{fig:many_dead}, we show full performance data over samples of $n_\mathrm{config}=200$ dead-qubit configurations (using the resampling strategy specified in Sec.~\ref{sec:failure_model}) for failure rates $p_\mathrm{failure}=0.5\%, 1\%, 2\%$ (from top to bottom) for both MIA (left column) and Auger (right column) for patches of size $L = 7, 9$. The solid curves are the \emph{median} $p_\mathrm{logical}$ over the sampled configurations while the dark shaded regions represent the upper and lower quartiles~\footnote{When reporting statistics over samples of dead qubit configurations from the ensemble parameterized by $p_\mathrm{failure}$, we ignore the statistical uncertainty in the estimation of each $p_\mathrm{logical}$ itself and plot the median/quartiles of the median of the estimated posterior distribution for $p_\mathrm{logical}$ in Figs.~\ref{fig:many_dead} and \ref{fig:histograms}.}; the dashed lines are the corresponding dead-component-free ($n_\mathrm{dead}=0$) curves, where the light shaded regions now represent statistical uncertainty for the individual $p_\mathrm{logical}$ values. In all, the MIA scheme leads to tight and impressively well-performing logical failure rates compared to Auger, with the differences being more pronounced at higher $p_\mathrm{failure}$ where Auger demonstrates an extremely accelerated performance degradation at the considered values.

In Fig.~\ref{fig:histograms}, we dive deeper into the performance distribution data of Fig.~\ref{fig:many_dead}, focusing on the slice at physical error rate $p_\mathrm{physical} = 5 \times 10^{-4}$ and patch size $L=9$ for concreteness. In the top row, we show a histogram of the estimated $p_\mathrm{logical}$ over the sampled configurations, for both MIA (green) and Auger (red), where solid vertical lines denote the respective medians (vertical dotted line is the $n_\mathrm{dead} = 0$ performance). The relative separation at $p_\mathrm{failure}=1\%$ seems competitive with recently developed schemes in Refs.~\cite{LUCI,SnL}, although making direct apples-to-apples comparisons with those works is not possible with the present data---one would need to implement either those schemes in our 3aux surface code or the MIA scheme in the CNOT-based surface code. Note that the large isolated peak in values at $p_\mathrm{failure} = 2\%$ for the Auger data at the bin centered near $p_\mathrm{logical} = 10\%$ corresponds to cases where the effective fault distance is reduced all the way to one such that the protocols are no longer error correcting; this situation is also apparent in the quartile data for Auger at $p_\mathrm{failure} = 2\%$ in Fig.~\ref{fig:many_dead}.

In the bottom row of Fig.~\ref{fig:histograms}, we show corresponding scatter plots of $(p_\mathrm{logical}^\mathrm{Auger}, p_\mathrm{logical}^\mathrm{MIA})$ over all $n_\mathrm{config}=200$ configurations sampled and binned in the histograms above. The points are colored according to the total number of dead qubits, $n_\mathrm{dead}$, in the corresponding configuration. Those points on the diagonal correspond to configurations where only data qubits are dead (since the schemes result in the same circuits), while a finite number of (bulk) auxiliary qubits gives rise to MIA's rather dramatic advantage, which is most prominent for the cases where Auger fails to be error correcting at all (see ``band'' emerging at lower-right) in which case there is up to a four-to-five order of magnitude performance separation between the two schemes.

\begin{figure}
    \centering
    \includegraphics[width=0.7\linewidth]{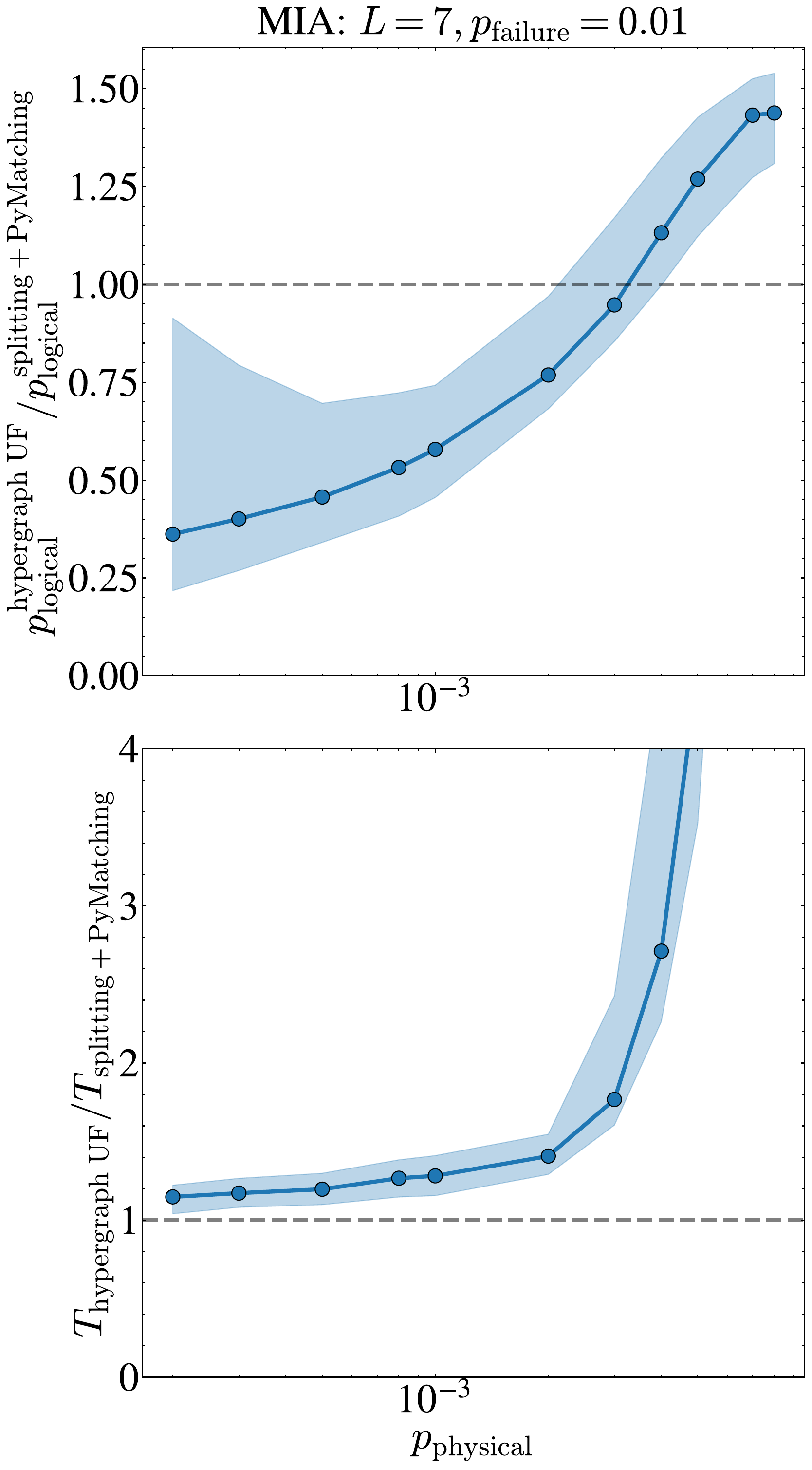}
    \caption{
    Relative logical error rate (top) and (simulation) runtime (bottom) versus $p_\mathrm{physical}$, comparing utilization of a weighted hypergraph UF decoder versus spitting + PyMatching for 3aux circuits under the MIA scheme. The shaded regions represent the middle 95\% of the obtained data over the same $n_\mathrm{configs}=200$ sampled configurations obtained for $p_\mathrm{failure}=1\%$ in Fig.~\ref{fig:many_dead}. Deep in the sub-threshold regime, we see a clear performance-runtime trade-off, where up to a factor of four in logical rate improvement is observed with a relatively minor ($\sim$20\%) increase in runtime. As above, we ignore statistical uncertainty in $p_\mathrm{logical}$ in these plots, which is at least partly the source of the increase in spread in the relative logical error rate at low $p_\mathrm{physical}$.
    }
    \label{fig:hgUF_comp}
\end{figure}

All of the data presented thus far in Figs.~\ref{fig:single_dead},\,\ref{fig:many_dead}, and \ref{fig:histograms} was obtained with PyMatching applied to a split hyperedge noise model using the check basis construction described in Sec.~\ref{sec:graph_realization}. Using the same check basis, we here consider decoders working directly with the decoding hypergraph, i.e., without needing to split the full circuit-level noise model (which can potentially lead to logical performance degradation). Here we investigate the putative logical performance / decoder runtime trade-off~\cite{delfosse_how_2023} in the context of our 3aux MIA circuits comparing the weighted hypergraph union-find (UF) decoder~\cite{delfosse_toward_2021} to our hyperedge splitting + PyMatching decoder~\footnote{We do not use a direct implementation of weighted hypergraph union-find, but instead a different algorithm that results in the same decoder outcome. Namely, we use the open-source minimum-weight parity factor (MWPF) decoder implementation from Ref.~\onlinecite{mwpf2025} and set the decoder timeout to zero.}. In Fig.~\ref{fig:hgUF_comp}, we show the relative logical error rates and runtimes~\footnote{The plotted quantity is the relative simulation time of the Monte Carlo sampling per shot, and thus the numerator and denominator each include a roughly constant, relatively small term corresponding to the time taken to propagate the sampled faults and compute the logical effects (whose syndromes are input to the decoder).} versus $p_\mathrm{physical}$ for the MIA scheme circuits corresponding to the same $n_\mathrm{configs} = 200$ dead-qubit configurations considered above for the $L=7$ 3aux system at $p_\mathrm{failure} = 1\%$. The shaded regions represent the middle 95\% of the obtained data over the sampled dead-qubit configurations, while the points represent the median. For the lowest $p_\mathrm{physical}$ considered, the median behavior corresponds to a factor of two to three improvement in logical error rate using weighted hypergraph UF compared to splitting + PyMatching. Interestingly, this improvement in logical error rate comes at the cost of only about a 20\% or less increase in runtime. For higher physical error rates, eventually weighted hypergraph UF gives a worse logical error rate than spitting + PyMatching, and the runtime increases rapidly (as expected, due to the poor scaling of the computational complexity for hypergraph UF).

In all, this data suggests that in the deep sub-threshold regime it may very well be worthwhile to consider such purely hypergraph-based decoders even for the surface code. Not only is the weighted hypergraph UF decoder performant and fast for these circuits, but it is of course also very generic.
Fully characterizing the performance-runtime trade-off across the landscape of decoder setups for this problem---including utilizing different hypergraph-based decoders such as heuristic modifications to hypergraph UF~\cite{Berent_2023} and relay-BP~\cite{muller_improved_2025}---is an interesting avenue for future work.

\section{Discussion and outlook}

\begin{figure*}
    \centering
    \includegraphics[width=0.9\linewidth]{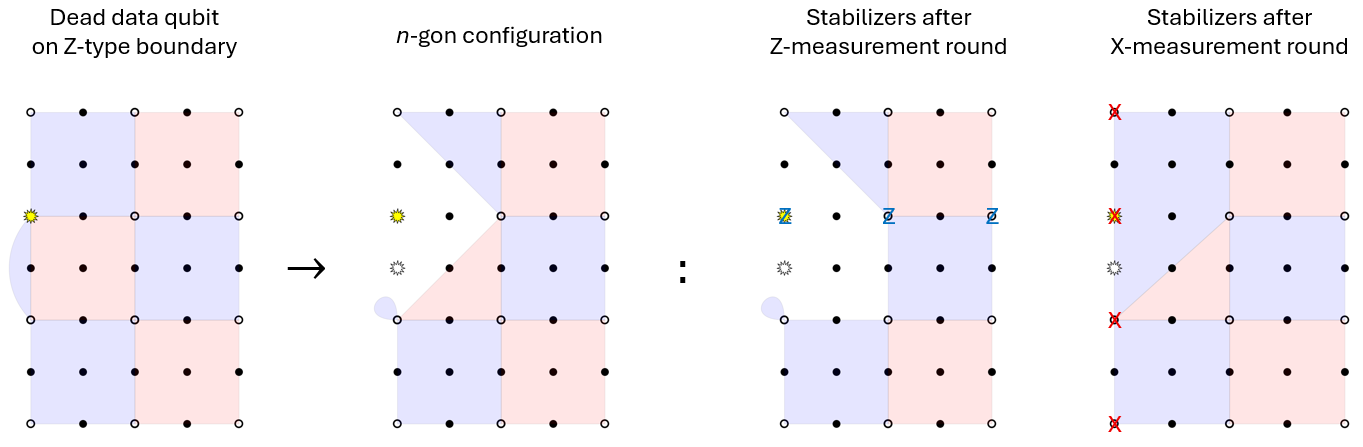}
    \caption{$n$-gons and instantaneous stabilizer groups resulting from a single dead data qubit on the boundary under the MIA scheme. A single dead data qubit on a $Z$-type boundary results in one $Z$-type 3-gon, one $X$-type 3-gon, and one $Z$-type 1-gon. This alters the instantaneous stabilizer group after completion of the $Z$- and $X$-measurement rounds as shown, thereby reducing the code distance by one for both logical observables (the components of representative logical strings that are missing relative to the dead-component-free case correspond to where the drawn representative logicals overlap the dead data qubit).}
    \label{fig:boundary}
\end{figure*}

The MIA scheme is a superstabilizer-based strategy for mitigating presence of failed components in a topological code such as the surface code. Here we have presented a decoding simulation pipeline and extensive simulations of the 3aux measurement-based surface code memory circuits under such a scheme. The MIA scheme is particularly simple and thereby appealing from a large-scale control perspective, yet maintains competitive logical performance. However, there are a few important points to call out. When performing \emph{logical operations} on a 3aux surface code (see, e.g., Refs.~\cite{bombin_logical_2023,gidney_inplace_2024,geher_error-corrected_2024,beverland2024faulttolerancestabilizerchannels}), we will not always be able to favorably orient the boundaries as in the memory protocol simulated here. Even if we use hook-preventing circuits~\cite{GransSamuelsson2024improvedpairwise}, there will seemingly be parts of the ``logical block'' for which the code distance is reduced by two for single auxiliary qubit failures, although only for a single logical observable (recall that Auger reduces the code distance by two for both observables). One would need to implement and simulate full logical operations in the presence of dead components to understand the importance of this effect quantitatively, which is beyond the scope of this paper. Nonetheless, the 3aux protocol and its baseline identity gate are indeed particularly well-suited for the MIA scheme and given the importance of logical idle in logical circuits for architectures with nearest-neighbor coupling~\cite{litinski_active_2022}, this perhaps bodes well. Still, it would be useful to explore more ``dynamic'' and sophisticated alterations to the circuits as in Refs.~\cite{LUCI,SnL} to further improve performance in the presence of dead auxiliary qubits during non-trivial logical operations.

While we have here focused on 3aux surface code circuits, it would be interesting to simulate the MIA scheme also for CNOT-based circuits. The separation between MIA and Auger will likely not be as pronounced as in 3aux as bulk single auxiliary qubit failure in a CNOT-based circuit amounts to replacing 4-qubit surface code stabilizers with measurement of the corresponding four 1-gons (i.e., direct readout of the data qubits; see Fig.~32 of Ref.~\onlinecite{GransSamuelsson2024improvedpairwise}); this gives rise to distance-two reduction in code distance, although for only one of the logical observables. Perhaps the more drastic improvement that would be harvested from employing the MIA is in its 
efficient measurement of superstabilizers in time, for either pipelined or interleaved schedules. It would indeed be worthwhile to pin down the quantitative effect of this scheduling consideration in CNOT-based architectures.

\acknowledgments

We thank Torsten Karzig, Rui Chao, Roman Lutchyn, Chetan Nayak, and Krysta Svore for useful discussions.

All circuits and corresponding data for the simulation results presented in this paper are available from a Zenodo repository~\cite{mishmash_2025_16748760}.

\appendix

\section{Boundary treatment} \label{app:boundary}

In this appendix, we briefly discuss how the $n$-gon-based MIA scheme allows the boundary to be treated on exactly the same footing as the bulk when adapting the surface code to the presence of dead components. In Fig.~\ref{fig:boundary}, we depict the situation encountered when a data qubit is dead along the boundary. The $n$-gon reduction rules~\cite{GransSamuelsson2024improvedpairwise} result in the affected two 4-gons and one 2-gon to be replaced by two 3-gons and one 1-gon. This effectively removes the 4-qubit $X$-type (red) stabilizer involving the dead qubit from the instantaneous stabilizer group after the $Z$-measurement round is complete, and heals the 4-qubit $Z$-type (blue) stabilizer involving the dead qubit into another 4-qubit $Z$-type stabilizer now involving the adjacent data qubit (that measured via the 1-gon) after the $X$-measurement round is complete. The net effect is reduction in the code distance (space-like instantaneous-stabilizer-group distance) by one for both logical observables (see Fig.~\ref{fig:boundary}), which is the same impact on code distance as single bulk data qubit death. The data in Fig.~\ref{fig:single_dead} explicitly shows logical performance data for single dead qubits along the boundary, and while it is difficult to see with the chosen color scale, boundary data qubit death is less costly than corresponding bulk death, due to entropic effects. Note that these considerations are not specific to the surface code implementation; however, they do require allowing 1-gons (mid-circuit readout of data qubits) in the set of available operations.

Similar considerations apply to dead auxiliary qubits along the boundary. Now specializing to 3aux, it is straightforward to show that dead auxiliary qubits in boundary 2-gons result in the code distance being reduced by one for only one logical observable. In our ``Auger'' simulations in this paper we treat qubit death for those qubits associated with boundary 2-gons (i.e., boundary data qubits and the 2-gon auxiliary qubits) using the MIA rules, for simplicity.

\bibliography{bibliography}

\end{document}